\documentclass{emulateapj}

\shorttitle{The Origin of Green Fuzzy Emission}
\shortauthors{De Buizer and Vacca}

\begin{document}

\title{Direct Spectroscopic Identification of the Origin of 'Green Fuzzy' Emission in Star Forming Regions}

\author{James M. De Buizer and William D. Vacca}
\affil{SOFIA-USRA, NASA Ames Research Center, MS 211-3, Moffett Field, CA 94035}

\begin{abstract}
``Green fuzzies'' or ``extended green objects'' were discovered in the recent \emph{Spitzer} GLIMPSE survey data. These extended sources have enhanced emission in the 4.5~$\mu$m IRAC channel images (which are generally assigned to be green when making 3-color RGB images from \emph{Spitzer} data). Green fuzzies are frequently found in the vicinities of massive young stellar objects, and it has been established that they are in some cases associated with outflows. Nevertheless, the spectral carrier(s) of this enhanced emission is still uncertain. Although it has been suggested that Br $\alpha$, H$_2$, [Fe II], and/or broad CO emission may be contributing to and enhancing the 4.5~$\mu$m flux from these objects, to date there have been no direct observations of the 4--5~$\mu$m spectra of these objects.

We report here on the first direct spectroscopic identification of the origin of the green fuzzy emission. We obtained spatially resolved $L$ and $M$ band spectra for two green fuzzy sources using NIRI on the Gemini North telescope. In the case of one source, G19.88-0.53, we detect three individual knots of green fuzzy emission around the source. The knots exhibit a pure molecular hydrogen line emission spectrum, with the 4.695~$\mu$m $\nu$ = 0--0 $S(9)$ line dominating the emission in the 4--5~$\mu$m wavelength range, and no detected continuum component. Our data for G19.88-0.53 prove that green fuzzy emission can be due primarily to emission lines of molecular hydrogen within the bandpass of the IRAC 4.5~$\mu$m channel.

However, the other target observed, G49.27-0.34, does not exhibit any line emission and appears to be an embedded massive young stellar object with a cometary UC HII region. We suggest that the effects of extinction in the 3--8~$\mu$m wavelength range and an exaggeration in the color stretch of the 4.5~$\mu$m filter in IRAC RGB images could lead to embedded sources such as this one falsely appearing ``green''.

\end{abstract}


\section{Introduction}

The \emph{Spitzer Space Telescope's} large scale survey of the Galactic plane, GLIMPSE (Galactic Legacy Infrared Mid-Plane Survey Extraordinaire; Benjamin et al. 2003) provides a rich archive of data relevant to the formation of stars at infrared wavelengths. This survey imaged the Galactic plane with the IRAC camera (Fazio et al.\ 2004) with filters at 3.6, 4.5, 5.8, and 8.0~$\mu$m at an angular resolution of $<$2$\arcsec$ in all bands. IRAC images of massive star formation regions often show conspicuous areas of extended emission that appear to have enhanced flux in the 4.5 $\mu$m filter (which is usually assigned to be the green channel when constructing 3-color images from \emph{Spitzer} IRAC data). These regions have therefore been given the name ``green fuzzies'' (Chambers et al. 2009) or ``extended green objects (EGOs)'' (Cyganowski et al. 2008).

The nature of these ``green" objects is not definitively known, but they appear to be largely associated with circumstellar gas and dust around, or outflows from, massive young stellar objects (MYSOs). However, since \emph{Spitzer} did not have spectroscopic capabilities covering the IRAC 4.5~$\mu$m filter, the specific spectral carrier(s) enhancing the emission within the wavelength range of just this one filter has not been conclusively determined.

The IRAC 4.5~$\mu$m filter (also known as ``Channel 2'') extends from 4.0 to 5.0~$\mu$m. If the enhanced emission arises from circumstellar material around the MYSO, lines such as Br~$\alpha$, Mg, and Ar, as well as the broader bands of deuterated PAHS at 4.48 and 4.65~$\mu$m (which have recently been found in ionized regions in the vicinity of massive stars; Peeters et al.\ 2004) could be the spectral carriers. However, if the emission arises from outflows or shocks, molecular and atomic shock indicators such as several H$_2$ and [Fe II] lines, as well as a broad CO ($\nu$ = 1--0) bandhead at 4.6~$\mu$m may be responsible for the enhanced ``green" appearance of these objects.

Several studies have tried to determine the spectral carriers responsible for the enhanced 4.5~$\mu$m emission via models or indirect observations. For instance, Noriega-Crespo et al.\ (2004) used IRS $5-37$~$\mu$m spectra to study the HH47 outflow (which has particularly strong 4.5~$\mu$m emission), and concluded based upon the H$_2$ line strengths at $\lambda>$5~$\mu$m that the green fuzzy emission may be due to H$_2$, with some contribution by small dust particles. Likewise, models by Smith \& Rosen (2005) predicted H$_2$ line intensities from outflows, and showed that a flux enhancement would be strongest in the 4.5~$\mu$m filter of IRAC. Alternatively, Marston et al.\ (2004) suggested that the IRAC 4.5~$\mu$m emission in the DR21 outflow is due to CO emission based on the spatial coincidence of green fuzzy features to features in 115 GHz CO maps.

We present here observations using Gemini Telescope's near-infrared spectrometer, NIRI, of $L$ and $M$ band spectra of two green fuzzies from the survey of Cyganowski et al.\ (2008). The wavelength range of our observations cover as much of the \emph{Spitzer} 3.6 and 4.5~$\mu$m filters as can be observed from the ground. From these observations, we attempt to conclusively identify the exact spectral carrier(s) responsible for green fuzzy emission.

\section{Observations}

We selected eight northern hemisphere sources of green fuzzy emission with high surface brightnesses from the survey of Cyganowski et al.\ (2008); however because of time constraints, only the two with the highest surface brightnesses were observed: G19.88-053 and G49.27-0.34. The observations were acquired with the NIRI instrument on Gemini in queue mode over 7 nights in 2009. The f/6 camera with a 6.4 pixel slit and a plate scale of 0.117$\arcsec$ pix$^{-1}$ was used for all observations. The $L$ and $M$ grisms were used to obtain spectra spanning the wavelength ranges $2.95 - 4.12$~$\mu$m and $4.47 - 5.75$~$\mu$m, respectively, with a spectral resolving power of $\sim$460 for both bands. All targets were observed in ``pair mode'', in which the objects were nodded along the slit between two positions (``A" and ``B" ). The images presented by Cyganowski et al.\ (2008) indicate that the emission from G19.88-053 and G49.27-0.34 have a spatial extents of $\sim$30$\arcsec$ and $\sim$20$\arcsec$, respectively; the separations between the two nod positions were therefore chosen to be 40$\arcsec$ and 30$\arcsec$, respectively, to account for their extended nature and ensure that extended emission was not subtracted off when the A-B pairs were subtracted from one another. In both cases, the slit was intended to be set up on the MYSO and rotated to cut across the brightest parts of the extended 4.5 $\mu$m emission. The slit position angles were 89$^{\circ}$ in the case of G19.88-0.53, and 295$^{\circ}$ for G49.27-0.34. However, because neither the MYSO nor the green fuzzy emission knots were visible in the $J$ band acquisition images used to set up the observations, a blind offset from a star $\sim$15$\arcsec$ away was required to position the slit on the target (NIRI cannot perform spectroscopic setups in the $M$ band). Inaccuracies in the absolute pointing and slit positioning could lead to the sources not being well-centered in the slit and (along with seeing fluctuations) could create significant variation in the final fluxes observed. Underscoring this point are the $M$ band observations of G19.88-0.53, which were taken over 5 nights and showed $\sim$30\% rms variation in the measured flux levels.

All observations at $L$ band were taken with frame times of 1.0 s, and the $M$ band observations were taken with frame times of 0.15 s for G49.27-0.34 and 0.2 s for G19.88-0.53. Coadding between 30 and 100 frames and between 25 and 100 A-B pairs was used to reach the total integration times. For G49.27-0.34 the total integration times were 1500 s at $L$ and 1380 s at $M$, with all data collected on the same night. The total integration time for G19.88-0.53 at $L$ was 1500 s, with all data obtained on a single night. At $M$, data were taken over 5 nights, with a final total exposure time of 6110 s.

Observations of a telluric standard star and flat field frames were obtained in the appropriate passbands on each night immediately before or after those of the target object. From those standard star observations the image quality of the observations was determined. In the case of G49.27-0.34, the $L$ and $M$ band images had a average full width at half maximum (FWHM) of 0.69$\arcsec$ and 0.71$\arcsec$, respectively. The $L$ band image for G19.88-0.53 had a FWHM of 0.61$\arcsec$, and the combined data from the 5 nights of $M$ band observations had an effective FWHM of 0.51$\arcsec$.

The data were reduced in IRAF using the Gemini reduction package specifically developed for reducing NIRI and GNIRS slit spectra. The processing steps included generating flat field images, generating ``sky" frames for each source frame from the observations at the opposite nod position, subtracting the sky frame from each image and dividing by the normalized flat field, stacking the sky-subtracted, flat-fielded frames and combining them. Residual sky emission was then subtracted from the result. The final two-dimensional spectral images (flux as a function of wavelength and position along the slit) for the two sources are shown in Figs. 1 and 2. The spectrum of the central source in each combined two-dimensional image was then traced and extracted. Multiple extraction apertures were defined for the various emission regions (including the MYSO itself) detected along the slit in the case of G19.88-053. For G49.27-0.34, two apertures were defined corresponding to the MYSO and the weak extended ``green'' emission region. Wavelength calibration was performed using sky emission lines.

Telluric corrections and flux calibration were performed in IDL using the spectra of the telluric standards observed on each night and the general version of the {\tt xtellcor} routine described by Vacca et al.\ (2003). For each of the various apertures defined for G19.88-0.53 the telluric-corrected $M$ band spectra from the separate observing nights were scaled to a common level (the scale values were derived from the observed and median strengths of the 4.69 $\mu$m emission line from the various nights) and median combined. The final $L$ and $M$ band spectra for these apertures as well as for the MYSO in this source are shown in Fig.\ 3. The final spectra for G49.27-0.34 are shown in Fig.\ 4. Line fluxes were measured on the green fuzzy spectra using IDL to directly integrate the flux under the features between two wavelength limits.

\section{Discussion}

\subsection{G19.88-0.53}

The extended green fuzzy emission from G19.88-0.53  morphologically appears to be an outflow from a MYSO (Fig.\ 1). We detect emission from three green fuzzy knots (GF1, GF2, and GF3) and from the MYSO itself. As far as we are aware, this is the first green fuzzy emission source to be directly observed spectroscopically at the wavelengths covered by the \emph{Spitzer} 4.5~$\mu$m channel. As shown in Fig.\ 3, we find conclusively that the main source of the enhanced green fuzzy emission in this source is the $\nu=0-0$ S(9) line of H$_2$ at 4.695~$\mu$m. We do not detect any other spectral features within the IRAC Channel 2 passband accessible from the ground; in particular we see no emission from the putative CO feature at 4.6~$\mu$m. Furthermore, we detect no continuum emission from the green fuzzy knots. These observations of G19.88-0.53 demonstrate that green fuzzy emission can indeed be created in outflows predominantly as a result of H$_2$ line emission, without the need for CO, [Fe~II], or Br $\alpha$ emission.

Interestingly, at the spatial resolution of our observations ($\sim$0.7$\arcsec$), we detect very little spatially extended H$_2$ emission along the slit in our two-dimensional spectral images of the outflow. Any possibly spatially extended H$_2$ emission must be either below our detection threshold, or located outside of the region sampled by our slit. This is consistent with the picture that H$_2$ emission is enhanced in the tips and wakes of bow shocks formed in outflows (Allen \& Burton 1993).

For GF1, the position of the peak seen in the NIRI spectral image is spatially coincident with the peak pixel seen in the 4.5 $\mu$m IRAC image. GF1 appears to have a lower surface brightness extension to the north that is below our sensitivity limit. The precise position of the GF2 peak cannot be distinguished in the IRAC image because of the poor IRAC resolution and the proximity of GF2 to the MYSO. The brightest pixel near GF3 in the IRAC image is actually coincident with a background star seen as a continuum source in our two-dimensional spectral image, but the extended 4.5 $\mu$m emission in the IRAC image does encompass the location of the knot detected in our NIRI spectral image.  The spatial coincidence between the knots detected in our slit spectra and the locations of the green fuzzy emission seen in the IRAC images adds further support to the conclusion that the H$_2$ emission we are seeing is responsible for the majority of the enhanced 4.5 $\mu$m emission from this source.

We attempted to extract flux densities of the individual green fuzzy knots from the \emph{Spitzer} IRAC data. However, because of the coarse resolution of the IRAC data and the fact that G19.88-0.53 lies partially in a ridge of dust that becomes apparent only at wavelengths $>$5 $\mu$m, aperture photometry could reliably be performed only on the GF1 knot.
We used several combinations of aperture radii and sky annuli for which aperture corrections are available in the IRAC Handbook\footnote{ http://ssc.spitzer.caltech.edu/irac/iracinstrumenthandbook/32/\#\_Toc257619125}. First we used a 2 pixel radius aperture centered on the source, with sky background measurement derived from the statistics in an annulus ranging from 2-6 pixels. We repeated these measurements with a 2 pixel aperture radius with a 12-20 pixel background annulus, a 3 pixel radius with a 3-7 pixel annulus, and a 3 pixel radius with a 12-20 pixel annulus. The average of these four values and their standard deviations were calculated.

Resultant point source flux densities for GF1 are 5.6$\pm$1.7, 18.9$\pm$6.8, and 15.8$\pm$6.3 mJy at 3.6, 4.5, and 5.8 $\mu$m, respectively. Because the diffuse, spatially variable background dust emission is much brighter at longer wavlengths, the 4.5 and 5.8 $\mu$m flux densities have a large uncertainties. In fact, there is no source visible at the location of GF1 in the IRAC 8.0~$\mu$m image above this background. As mentioned above, GF2 is too close to the MYSO and GF3 lies $\sim$2.5$\arcsec$ from a background star, so reliable flux density estimates for these two sources could not be obtained. The background star near GF3 can be seen in 2MASS images at $J, H$ and $K$, and demonstrates the need for observations with higher spatial resolution than afforded by \emph{Spitzer} in isolating the flux from the green fuzzy knots themselves.

Is the emission we are seeing from H$_2$ lines enough to account for the enhancement seen in the 4.5~$\mu$m IRAC image? As stated above, the extent of the green fuzzy emission seen in the IRAC images is clearly far larger than the width of our slit, and we are likely only detecting only the brighter knots of emission. With these caveats in mind, we converted the above IRAC photometry to surface brightnesses and compared them to the surface brightnesses derived from our NIRI spectra. The above IRAC flux densities translate to surface brightnesses of 4.8($\pm$1.2)$\times$10$^{-17}$ Wm$^{-2}$$\mu$m$^{-1}$arcsec$^{-1}$ at 3.6~$\mu$m, and 1.0($\pm$0.3)$\times$10$^{-16}$ Wm$^{-2}$$\mu$m$^{-1}$arcsec$^{-1}$ at 4.5~$\mu$m. From the NIRI spectra we derive surface brightnesses of 3.2($\pm$0.2)$\times$10$^{-17}$ Wm$^{-2}$$\mu$m$^{-1}$arcsec$^{-1}$ within the 3.6~$\mu$m IRAC band, and 6.2($\pm$0.4)$\times$10$^{-17}$ Wm$^{-2}$$\mu$m$^{-1}$arcsec$^{-1}$ within the 4.5~$\mu$m IRAC band. This latter value is derived by taking into account the estimates for the fluxes of the unobserved 4.180 $\mu$m $\nu$ = 0--0 $S(11)$ and the 4.408 $\mu$m $\nu$ = 0--0 $S(10)$ lines at the measured gas temperature and the measured ortho-to-para ratio of 1.55 (both are derived in $\S$3.1.1). These data show that the H$_2$ lines observed in the NIRI spectra contribute $\sim$66\% of the surface brightness measured in the IRAC 3.6~$\mu$m image and $\sim$62\% of the surface brightness in the IRAC 4.5~$\mu$m image.

We converted these surface brightnesses to flux densities by scaling by the size of the NIRI extraction aperture. This yields flux densities of 7.1($\pm$2.3)$\times$10$^{-17}$ Wm$^{-2}$$\mu$m$^{-1}$ at 3.6~$\mu$m, and 1.2($\pm$0.4)$\times$10$^{-16}$ Wm$^{-2}$$\mu$m$^{-1}$ at 4.5~$\mu$m. The flux densities measured from the H$_2$ lines in our spectra are 4.7($\pm$0.3)$\times$10$^{-17}$ Wm$^{-2}$$\mu$m$^{-1}$ for the 3.6~$\mu$m IRAC bandpass, and  7.5($\pm$0.5)$\times$10$^{-17}$ Wm$^{-2}$$\mu$m$^{-1}$ for the 4.5~$\mu$m IRAC bandpass. The latter value again takes into account the missing 4.180 and 4.408~$\mu$m lines and the measured ortho-to-para ratio.


From these comparisons it is reasonable to conclude that the majority of the observed IRAC flux is due to the H$_2$ lines seen in our spectra. Moreover, given the errors in both the surface brightnesses and flux densities derived above, we can conclude that it is possible that \emph{all} of the green fuzzy emission is due to H$_2$ lines to within our measurement errors (and this does not even take into account the fact that the spectral lines are the median flux values from across all observations which varied on the order of 30\% from night to night; see $\S$2).

\subsubsection{The physical conditions of the H$_2$ emitting gas in G19.88-0.53}

Fluxes for the H$_2$ emission lines seen in the $L$ and $M$ band spectra of the three green fuzzies (GF1, GF2, and GF3) detected in our observations of G19.88-053 are given in Table 1. The absence of the 3.376~$\mu$m $\nu$ = 4--3 $O(3)$, 3.437~$\mu$m $\nu$ = 2--1 $O(5)$, and 3.662~$\mu$m $\nu$ = 3--2 $O(5)$ lines in our spectra indicates that the H$_2$ excitation mechanism is due to shocks and not UV fluorescence. The fluorescent H$_2$ models of Black \& van Dishoeck (1987) indicate that these three lines should have strengths $\sim$$30-60$\% of that of the 3.234~$\mu$m line. Given the signal-to-noise of our spectra, we can confidently rule out any substantial contributions due to fluorescent excitation of H$_2$.

As we detect several H$_2$ lines from GF1, we can derive a gas temperature from the H$_2$ excitation diagram (see e.g., Rosenthal et al.\ 2000). Assuming the lines are optically thin, we derived column densities for each upper level from
$$N_{\rm upper} = \frac{4 \pi \lambda I_{\rm obs}}{h c A}$$
where $I_{\rm obs}$ is the observed line intensity (W m$^{-2}$ sr$^{-1}$) and $A$ is the radiative transition probability for each line. The line intensities were computed from the measured line fluxes and the spatial areas corresponding to the spectral extractions. The excitation temperature of the gas can be derived by plotting the values of the estimated column densities divided by the level degeneracies $g$ as a function of the energy of the upper level $E_{upper}$ (see Fig.\ 5). Although we detected fewer H$_2$ lines for the other two green fuzzy knots, we included the estimated column densities from the available transitions from these regions in Fig.\ 5. As can be seen from the excitation diagram, the results for each of the three green fuzzies are consistent with a linear relation between $\log N_{\rm upper}/g$ and $E_{upper}$. The slope of the line is inversely proportional to the excitation temperature, while the intercept yields an estimate of the total H$_2$ column density, from the relation
$$ \ln \Bigl(\frac{N_{upper}}{g}\Bigr) = \ln \Bigl[\frac{N_{H_2}}{Z(T_{excit})}\Bigr] - \Bigl(\frac{E_{upper}}{kT_{excit}}\Bigr)~~~,$$
where $N_{H_2}$ is the total H$_2$ column density and $Z(T_{excit})$ is the H$_2$ partition function, computed using the expressions given by Irwin (1987). From a linear fit to the data shown in Fig.\ 4 for each green fuzzy we find $T_{excit} = 2570 \pm 70$ K and $N_{H_2} = 9.7 (\pm 1.0) \times 10^{17}$ cm$^{-2}$ for GF1, $T_{excit} = 2100 \pm 180$ K and $N_{H_2} = 4.0 (\pm 1.7) \times 10^{17}$ cm$^{-2}$ for GF2, and $T_{excit} = 2520 \pm 80$ and $N_{H_2} = 4.6 (\pm 0.6) \times 10^{17}$ cm$^{-2}$ K for GF3. If all three knots of emission are excited by the energetics of a single outflow from the MYSO, then the similarity of all three temperature values seems reasonable. We note that the column densities are strictly lower limits as we have not made any corrections for (unknown) extinction along the line of sight to the knots.

We note that one line in Fig.\ 4, the $\nu = 0 - 0$ S(8) para-H$_2$ line at 5.052~$\mu$m, seems to be too strong compared to the linear, single excitation temperature relation derived (primarily) from the ortho-H$_2$ lines.
(The other para-H$_2$ line in our spectra is the very weak $\nu = 0 - 0$ S(12) line at 3.996~$\mu$m, whose line flux is highly uncertain.) Following Puxley et al.\ (2000), we used the relative strengths of the well-detected ortho-H$_2$ 4.695~$\mu$m line and the 5.052~$\mu$m line to estimate the ortho-to-para ratio from
$$ \gamma = \frac{I_o \lambda_o}{I_p \lambda_p}\frac{A_p (2J_p + 1)}{A_o (2J_o + 1)} \exp\Bigl[{-\Bigl(\frac{E_p - E_o}{kT_{excit}}\Bigr)}\Bigr]$$
where subscript $o$ refers to the ortho transition while $p$ refers to the para transition. From the observed line fluxes we estimate the ortho-to-para ratio to be $1.55 \pm 0.11$.

The relative line fluxes we find are in reasonable agreement with those predicted by the model of Kaufman \& Neufeld (1996) for a shock velocity of 40 km s$^{-1}$ and a pre-shock hydrogen density of $10^6$ cm$^{-3}$, after taking into account the possible enhancement of the 5.052~$\mu$m line due to the low value of $\gamma$. The relative fluxes for the $L$ band lines are also in reasonable agreement with those given by Ybarra \& Lada (2009) for a shock model with $2000 < T_{\rm gas} < 4000$ K.

However, our ortho-to-para ratio value of 1.55 seems anomalously low for the gas temperature we have derived. Observations of H$_2$ lines in HH54 by Neufeld et al.\ (2006) have shown that values of 1.55 are not uncommon in observations of outflows, but such low ortho-to-para ratios are generally found for much lower gas temperatures ($\sim$1100~K). Models by Neufeld et al.\ (2006) show that for our derived temperature of $\sim$2600~K, we should find an ortho-to-para ratio of $\sim$3. Although we cannot explain our low value, it should be noted that for HH54, Ybarra \& Lada (2009) calculate gas temperatures of 2000-3300~K from IRAC photometry and modeling the H$_2$ line contributions within the IRAC filters.

\subsubsection{The nature of the MYSO in G19.88-0.53}

The spectrum of the MYSO powering the G19.88-0.53 outflow is shown in Fig.\ 3. The presence of a deep absorption feature attributed to solid-phase carbon monoxide at 4.67~$\mu$m indicates that there is at least 20 mags of visual extinction toward this source (Shuping et al.\ 2000).
In addition to CO absorption, we detect carbonyl sulphide (OCS; Palumbo, Tielens, \& Tokunaga 1995) in absorption at 4.9~$\mu$m, as well as the ``XCN'' absorption feature at 4.62~$\mu$m (Pendleton et al.\ 1999). These absorption features are seen in the spectra of very embedded MYSOs (e.g.,  NGC7438 IRS9; Gibb et al.\ 2004) and their presence is consistent with the lack of detectable flux in our spectrum at $\lambda$ $<$ 3.5~$\mu$m due to high extinction. Also present in our $L$ band spectrum are shallow and wide absorption features due to methanol, centered at 3.53 and 3.95~$\mu$m, which have also been seen towards luminous MYSOs (Dartois et al.\ 1999).

Although the MYSO appears to be greenish in Fig.\ 1, we do not detect any emission lines in its spectrum. All indications are that this is simply a deeply embedded source. This suggests that, in addition to being extinguished at $\lambda$ $<$ 3.5~$\mu$m, this source is likely to have a deep silicate absorption feature at 9.7~$\mu$m (see Fig.\ 12 of Dopita et al.\ 2006). The presence of a deep silicate absorption feature will depress the flux in the 8~$\mu$m IRAC image that was used by Cyganowski et al.\ (2008) as the red channel in their RGB composite images. A suppression of emission due to extinction/absorption within both the IRAC 3.5 and 8.0~$\mu$m bands may, in part, be responsible for the greenish appearance of the MYSO itself (this will be further discussed in $\S$3.3).

Comparing the point-source photometry from the IRAC images to the flux levels of the NIRI spectra shows that we recover only approximately one sixth of the overall flux from the MYSO in our spectra. The MYSO certainly looks point-like in our spectral images, yet the IRAC images do appear to show much extended emission nearby. Under the assumption that the source is resolved in the \emph{Spitzer} IRAC images, we are able to recover the flux densities of the source in the same manner as for the green fuzzy, GF1; the derived flux densities for the MYSO are are 10.0($\pm$1.1)$\times$10$^{-15}$ Wm$^{-2}$$\mu$m$^{-1}$ at 3.6~$\mu$m, and 2.1($\pm$0.2)$\times$10$^{-15}$ Wm$^{-2}$$\mu$m$^{-1}$ at 4.5~$\mu$m in an area equivalent to the NIRI slit aperture. Those values are overplotted on the spectra in Fig.\ 3. There are a few reasons that the total point source flux is so low in our spectra compared to that in the IRAC images, while at the same time we recover the surface brightnesses derived from the IRAC images. First, it may be very likely given the blind pointing described in $\S$2 that we were not set up on the peak of the MYSO in the slit. Second, it is likely that the source is not a point source, but has some extended component that constitutes a significant part of the measured overall flux from the MYSO in the IRAC images. We suggest that a combination of these two influences may be at work, but are unable to quantify exactly how much each contributes.

\subsection{G49.27-0.34}

G49.47-0.34 is another high surface brightness green fuzzy. However, unlike G19.88-0.53, this source does not morphologically look like an outflow. Instead it is very similar in appearance to a cometary-shaped ultracompact HII (UC HII) region (Fig.\ 2).

Although our total $M$ band exposure time for this source was significantly smaller than that for G19.88-0.53, the time for a single nightly data set was the same as that for G19.88-0.53.  Nevertheless, we detect no emission from H$_2$ or any other lines in our spectra of G49.27-0.34 whereas the H$_2$ lines were readily apparent in the individual nightly spectra of G19.88-0.53. The spectrum of the extended emission of the source appears to be due to the dust continuum from the UC HII region. Again, the presence of a deep CO absorption feature in the MYSO indicates that this source is heavily extinguished.

Given that the emission in this source is extended where the putative green fuzzy is located, we can employ the same technique as for GF1 in G19.88-0.53 and calculate the flux densities from the IRAC data and compare them directly to our spectra. We derive from the IRAC data the following flux densities within an area equivalent to the NIRI aperture used: 3.09($\pm$0.02)$\times$10$^{-16}$ Wm$^{-2}$$\mu$m$^{-1}$ for the 3.6 $\mu$m IRAC bandpass, and 9.55($\pm$0.98)$\times$10$^{-16}$ Wm$^{-2}$$\mu$m$^{-1}$ for the 4.5 $\mu$m IRAC bandpass. These values are plotted on the spectra of the green fuzzy in Fig.\ 5, and show that the continuum emission detected in our spectra represents all of the flux from the putative green fuzzy emission.

A similar calculation was done for the MYSO itself in G49.27-0.34. Like the MYSO in G19.88-0.53, the derived point source fluxes from the IRAC images are much brighter than our spectra, indicating that we are likely not dealing with a point source. Therefore, we derived from the IRAC data the following flux densities from within an area equivalent to the NIRI apertures used: 2.15($\pm$0.48)$\times$10$^{-15}$ Wm$^{-2}$$\mu$m$^{-1}$ for the 3.6 $\mu$m IRAC bandpass, and 11.2($\pm$3.0)$\times$10$^{-15}$ Wm$^{-2}$$\mu$m$^{-1}$ for the 4.5 $\mu$m IRAC bandpass. These values are plotted on the spectra in Fig.\ 5, and show that the we recover the IRAC flux from the extended dust continuum of both the MYSO and putative green fuzzy in our spectra.

\subsection{Why might compact embedded YSOs appear green?}

In the survey of Cyganowski et al.\ (2008), 3-color images of the IRAC data were produced using the 3.6, 4.5 and 8.0~$\mu$m data. As discussed in $\S$ 3.1.2, dust extinction in highly embedded regions can depress the 3.6 and 8.0~$\mu$m emission from sources. Indebetouw et al. (2005) also demonstrated that the Galactic dust extinction law actually flattens from 5--8~$\mu$m, rather than decreasing as one might expect. They also emphasized that the 9.7~$\mu$m absorption feature can affect the flux observed in the IRAC 8 $\mu$m filter.

Another probable source of artificially increased 4.5~$\mu$m flux is discussed by Povich et al.\ (2007). They claim that many of the 4.5~$\mu$m ``extended sources'' can appear to be ``green'' solely due to an exaggeration in the color stretch used. Since the 4.5~$\mu$m filter does not encompass any PAH emission features, images in this filter tend to have much fainter diffuse emission than those in the 5.8 or 8.0~$\mu$m bands, and to compensate for this the stretch is often changed to brighten the emission relative to that of the other bands for display purposes.

The compact MYSO in G19.88-0.53 may appear artificially green in the 3-color images of Cyganowski et al.\ (2008) due to a combination of these aforementioned effects. Since UC HII regions on average lie behind 30 magnitudes of extinction or more (Dopita et al.\ 2006), it is possible that for these same reasons even a source like G49.47-0.34 may appear as an extended but artificially ``green'' source.

\section{Conclusions}

We have obtained ground-based $L$ and $M$ band spectra of two high surface brightness green fuzzy objects from the sample of Cyganowski et al.\ (2008). Our observations indicate that for the outflow source G19.88-053, the green fuzzy emission is found within the outflow itself and is due primarily to H$_2$ emission lines within the IRAC 4.5 $\mu$m band arising from discrete, spatially compact knots. We detect no continuum emission, or any other emission features. Measurements of the H$_2$ lines in our spectra suggest that the knots comprise shock heated gas with temperatures of $2100-2600$ K and column densities on the order of $4-10 \times 10^{17}$ cm$^{-2}$. For one of the knots in this source, we find an unusually low H$_2$ ortho-to-para ratio of 1.55.

Our observations of G49.47-0.34 indicate that this source consists of a MYSO deeply embedded within an UC HII region. We see no sign of an outflow. Both the MYSO and the limb of the UC HII region appear ``green" but exhibit only bright continuum emission; we do not detect any emission lines within the IRAC 4.5~$\mu$m bandpass. As in the case of the MYSO in G19.88-0.53, we suggest that there may be reasons other than enhanced line emission for the ``green'' appearance of sources that are not outflows.

\acknowledgments Based on observations obtained at the Gemini Observatory, which is operated by the Association of Universities for Research in Astronomy, Inc., under a cooperative agreement with the NSF on behalf of the Gemini partnership: the National Science Foundation (United States), the Science and Technology Facilities Council (United Kingdom), the National Research Council (Canada), CONICYT (Chile), the Australian Research Council (Australia), Ministério da Ciência e Tecnologia (Brazil) and Ministerio de Ciencia, Tecnología e Innovación Productiva (Argentina). Gemini program identification number associated with the data is GN-2009B-Q-79. We would like to thank Andrew Stephens of Gemini Observatory for his help with aspects of the NIRI data reduction.

{\it Facilities:} \facility{Gemini:Gillett (NIRI)}

\clearpage
\begin{deluxetable}{lcccccccccc}
\tabletypesize{\footnotesize}
\tablewidth{0pt}
\tablecaption{Observed H$_2$ Line Fluxes for
G19.88-0.53 Green Fuzzies}
\tablehead{	
\colhead{} &\colhead{} &\colhead{}&\multicolumn{2}{c}{GF1} &\colhead{} &\multicolumn{2}{c}{GF2} &\colhead{} &\multicolumn{2}{c}{GF3}\\	
\cline{4-5} \cline{7-8} \cline{10-11}	\\	
\colhead{Transition} &\colhead{$\lambda$} &\colhead{}	 &\colhead{Line Flux}	  &\colhead{Error}&\colhead{} &\colhead{Line Flux}	  &\colhead{Error} &\colhead{}	 &\colhead{Line Flux}	  &\colhead{Error}\\
\colhead{} &\colhead{($\mu$m)}	&\colhead{} &\colhead{(W m$^{-2}$)} &\colhead{(W m$^{-2}$)}	&\colhead{} &\colhead{(W m$^{-2}$)} &\colhead{(W m$^{-2}$)} &\colhead{} &\colhead{(W m$^{-2}$)} &\colhead{(W m$^{-2}$)}
}
\startdata	
$\nu$=1--0 $O(5)$     &	3.234	& &	1.92E-17	&	1.02E-18  & &	\nodata	    &	\nodata	& &	9.20E-18	&	5.28E-19	\\
$\nu$=1--0 $O(7)$	    &	3.807	& &	8.32E-18	&	 1.10E-18	& &	\nodata	    &	\nodata	& &	\nodata	&	 \nodata	\\
$\nu$=0--0 $S(13)$	    &	3.847	& &	1.12E-17	&	 1.24E-18	& &	1.54E-18	    &	5.14E-19	& &	4.74E-18	 &	 5.67E-19	 \\
$\nu$=0--0 $S(12)$	    &	3.996	& &	4.97E-18	&	 9.87E-19	& &	\nodata	    &	\nodata	& &	\nodata	&	 \nodata	\\
$\nu$=0--0 $S(9)$	    &	4.695	& &	3.13E-17	&	 1.23E-18	& &	7.42E-18	    &	4.87E-19	& &	1.02E-17	 &	 6.16E-19	 \\
$\nu$=0--0 $S(8)$	    &	5.052	& &	2.10E-17	&	 1.18E-18	& &	2.86E-18      &	3.79E-19	& &	3.44E-18	&	 5.01E-19	
\enddata
\tablecomments{The 4.695 and 5.052~$\mu$m line fluxes were measured on a median spectrum generated from six M band data sets taken over 5 nights. The L band measurements are from a single data set taken on one night. Errors are purely statistical. Absolute errors are expected to be on the order of $\sim$30\%.}
\end{deluxetable}

\clearpage

\begin{figure}
\includegraphics[scale=.74,angle=0]{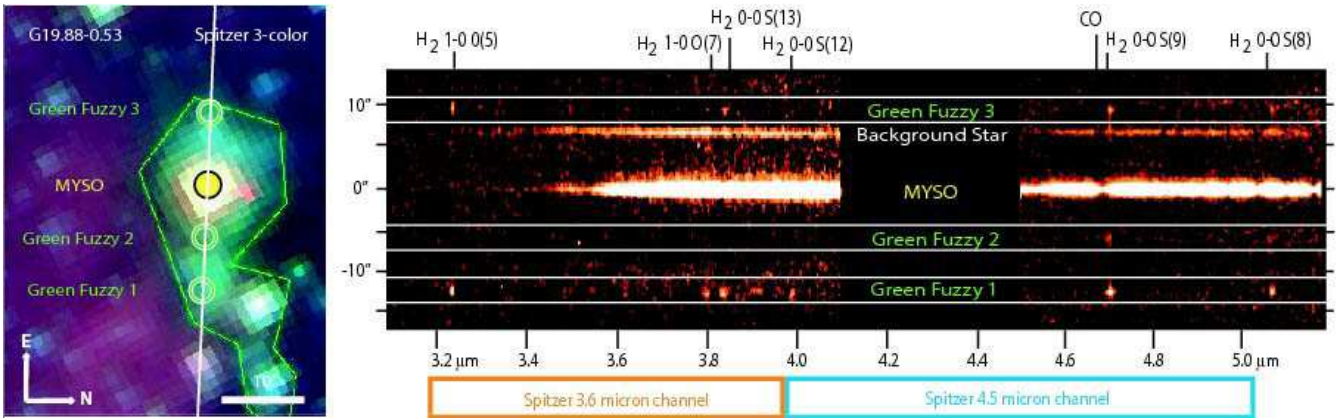}
\caption{Green fuzzy emission from G19.88-0.53. (Left) A 3-color composite image from Cyganowski et al.\ (2008), with 3.6~$\mu$m as blue, 4.5~$\mu$m as green, and 8.0~$\mu$m as red. The area containing extended green emission is encompassed in the green contour. The NIRI slit position and angle used for spectroscopy is shown by the white line. (Right) 2-D images of the combined L and M band spectra (with 3-pixel smoothing for display purposes only) are shown with the same vertical spatial scaling as the figure on the left. White lines encompass the spectra of the three green fuzzy knots detected. The bandpasses of the IRAC 3.6 and 4.5~$\mu$m filters are shown at the bottom. The only detectable emission from the three green fuzzies are lines of shock excited H$_2$, which are labeled at the top. The absorption feature of CO is also marked. No data are present from 4.10 to 4.55~$\mu$m where the atmosphere is opaque.}
\end{figure}


\begin{figure}
\includegraphics[scale=0.74,angle=0]{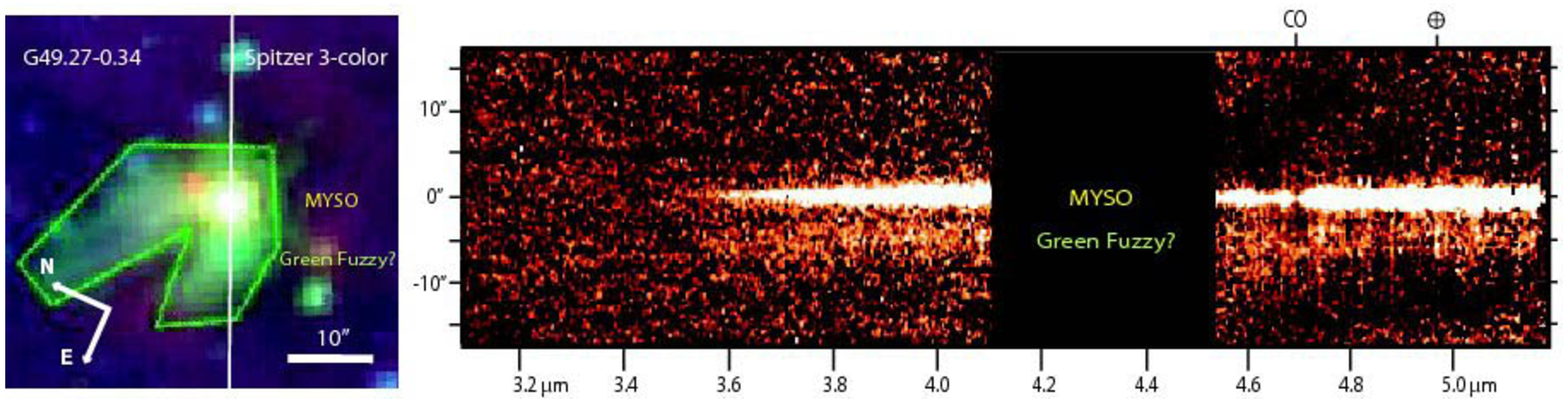}
\caption{Green fuzzy emission from G49.27-0.34. (Left) A 3-color composite image from Cyganowski et al.\ (2008), with 3.6~$\mu$m as blue, 4.5~$\mu$m as green, and 8.0~$\mu$m as red. The area containing extended green emission is encompassed in the green contour. The NIRI slit position and angle used for spectroscopy is shown by the white line. (Right) 2-D images of the combined L and M band spectra with 3-pixel smoothing for display purposes only) are shown with the same vertical spatial scaling as the figure on the left. The only detectable emission from the extended ``green'' emission area is continuum. The absorption feature of CO is marked, as well as a region that results from poor sky subtraction (encircled cross symbol). No data are present from 4.10 to 4.55~$\mu$m where the atmosphere is opaque.}
\end{figure}

\clearpage

\begin{figure}
\includegraphics[scale=0.90,angle=0]{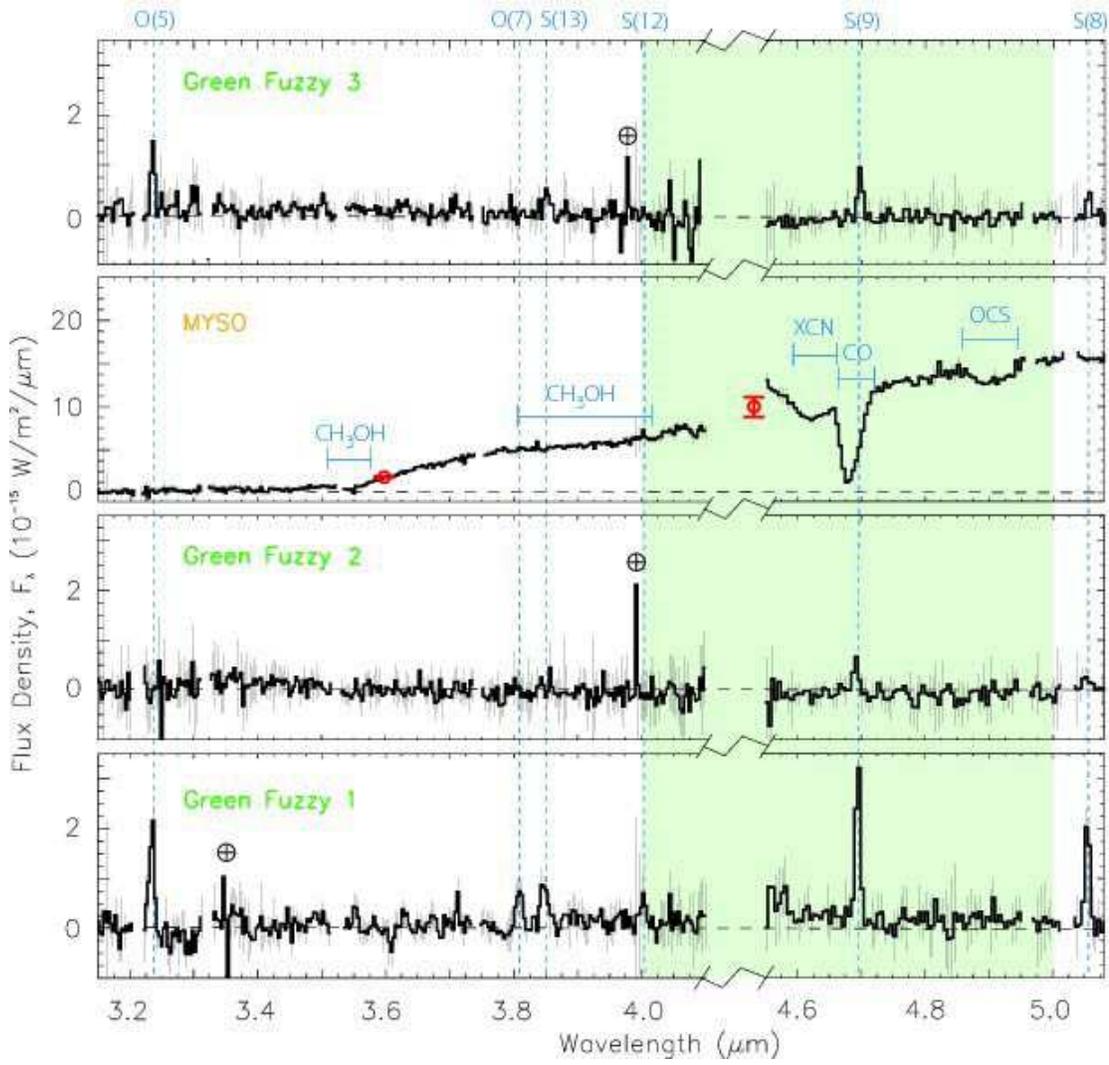}
\caption{Combined $L$ and $M$ band spectra extracted from the locations within G19.88-0.53 marked as circles in Figure 1.  The light green area delineates the passband of the IRAC 4.5~$\mu$m filter. The only detectable emission features from the three green fuzzies are lines of shock excited H$_2$, which are delineated by vertical dashed blue lines with the transitions labeled at the top. Areas of the spectra with missing data were removed because of poor sky subtraction. In some cases bad pixels create false features which are delineated by the encircled cross symbol. Flux densities derived from the IRAC image photometry of the MYSO are plotted as red circles. No data are present from 4.10 to 4.55~$\mu$m where the atmosphere is opaque.}
\end{figure}

\clearpage

\begin{figure}
\includegraphics[scale=0.90,angle=0]{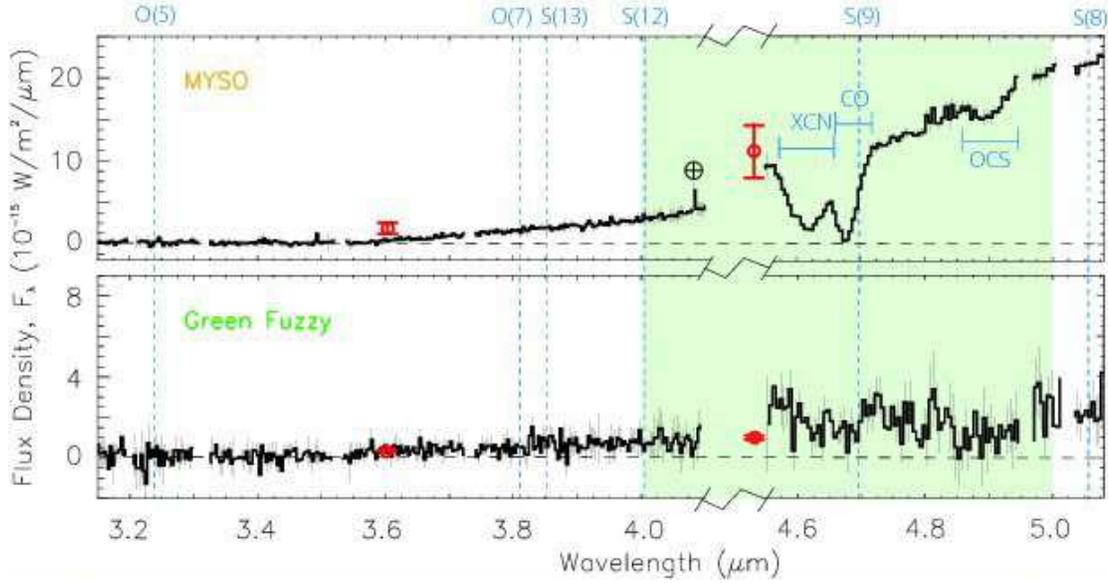}
\caption{Combined $L$ and $M$ band spectra for G49.27-0.34.
The light green area delineates the passband of the IRAC 4.5~$\mu$m filter. The only detectable emission features from the putative green fuzzy area is continuum emission. The expected lines of shock excited H$_2$, which are delineated by vertical dashed blue lines with the transitions labeled at the top. Areas of the spectra with missing data were removed because of poor sky subtraction. In some cases bad pixels create false features which are delineated by the encircled cross symbol. Flux densities derived from the IRAC image photometry of the MYSO and the putative green fuzzy are plotted as red circles. No data are present from 4.10 to 4.55~$\mu$m where the atmosphere is opaque.}
\end{figure}


\begin{figure}
\includegraphics[scale=0.90,angle=0]{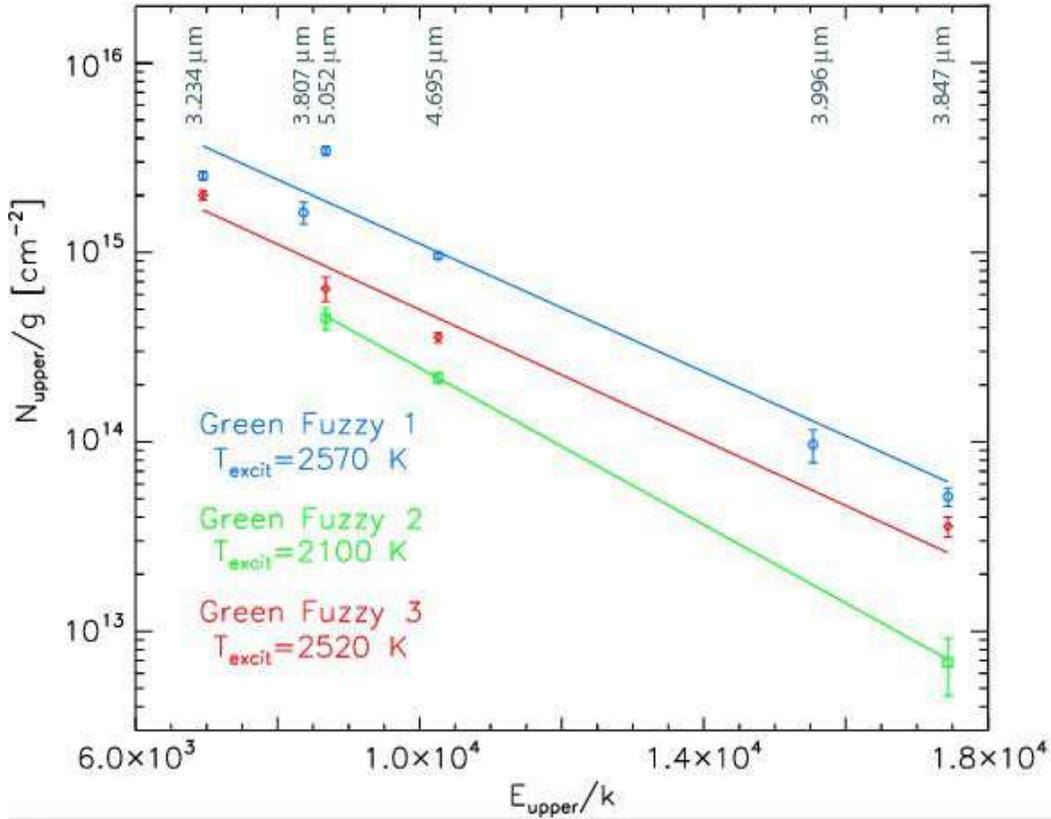}
\caption{Excitation diagram of the H$_2$ level column density distribution towards the green fuzzy knots in G19.88-0.53. We are able to fit these data with single temperatures value for each of the green fuzzy knots (solid lines).}
\end{figure}

\clearpage

\end{document}